\documentclass{article}  

\usepackage[dvipsnames,table,xcdraw]{xcolor}
\usepackage{soul}
\usepackage{amsmath}
\usepackage{mathtools}
\usepackage{amsfonts}
\usepackage{listings}
\usepackage{lstbayes}
\usepackage{booktabs}
\usepackage{graphicx}
\usepackage{geometry}
\usepackage{pgfplots}
\usepackage{caption}
\usepackage{subcaption}
\usepackage[authoryear]{natbib}
\usepackage{float}
\usepackage{url}
\usepackage{lineno}
\newif\ifproofread
\sethlcolor{Lavender}

\newcommand{\rev}[1]{%
\ifproofread
\hl{#1}%
\else
#1%
\fi
}

\graphicspath{{../}}

\DeclareMathOperator{\logit}{logit}

\DeclareMathOperator{\Normal}{Normal}

\title{GLM for partially pooled categorical predictors with a case study in biosecurity}
\author{Christopher M.\ Baker \and Howard Bondell \and Nathaniel Bloomfield \and Elena Tartaglia \and Andrew P.\ Robinson}
\begin{document}


\maketitle
\begin{abstract}
National governments use border information to efficiently manage the biosecurity risk presented by travel and commerce. In the Australian border biosecurity system,\rev{ data about cargo consignments are collected from records of directions: that is, the records of actions taken by the biosecurity regulator. This data collection is complicated by the way directions for a given entry are recorded. An entry is a collection of import lines} where each line is a single type of item or commodity. Analysis is simple when the data are recorded in line mode: the directions are recorded individually for each line. The challenge comes when data are recorded in container mode, because the same direction is recorded against each line in the entry. In other words, if at least one line in an entry has a non-compliant inspection result, then all lines in that entry are recorded as non-compliant. Therefore, container mode data creates a challenge for estimating the probability that certain items are non-compliant, because matching the records of non-compliance to the line information is impossible. We develop a statistical model to use container mode data to help inform biosecurity risk of items. We use asymptotic analysis to estimate the value of container mode data compared to line mode data, do a simulation study to verify that we can accurately estimate parameters in a large dataset, and we apply our methods to a real dataset, for which important information about the \rev{risk of non-compliance} is recovered using the new model. 
\end{abstract}

\section{Introduction}

Invasive species pose a multifaceted threat to society, leading to reductions in agricultural productivity, as well as damages to the environment, human health, and the economy \citep{kumar_rai_invasive_2020}. Considerable effort is devoted to managing invasive species \citep{jardine_estimating_2018}, in either eradicating them \citep{baker_recent_2020,holmes_globally_2019, wenger_estimating_2017,helmstedt_prioritizing_2016}, or suppressing their numbers to reduce damages \citep{binny_long-term_2021,brook_effects_2012,sharov_slow_2002}. The costs associated with managing invasive species provides governments with an incentive to manage biosecurity risks at national borders to prevent the establishment of the new species. 

Given the massive scale of global trade, biosecurity regulators need to be able to allocate their resources efficiently, but to do so they must understand the risks associated with various entries. At the border, one of the most effective ways of achieving this is by using the outcome of previously conducted inspections as intelligence that informs future operations. This allows regulators to identify high risk commodities and importers, and modify their inspection targets and policies accordingly.

However, gaining insight from border inspections is a massive logistical challenge, as data must be consistently recorded in a format that makes this analysis possible. Putting infrastructure into place to collect this data, and collecting it accurately can be expensive and challenging. Often, shortcuts may be taken that render the data less valuable in analysing patterns of risk.

In Australia systems have been put in place to capture biosecurity data since the early 90's; \rev{these data are extracted from the directions applied to the cargo in question. Directions are used to control the movement and direct the assessment and management of goods subject to biosecurity control, and the \emph{modes} discussed in this paper correspond to how those directions are carried out: at the line level, or at the entry level.}  As such, in line mode, the details of which items within an entry are inspected and found to be compliant or non-compliant are fully recorded. However, in container mode, the results of an inspection are applied to all items within an entry. Container mode was introduced so that containers could be released piecemeal in order to minimise the bottlenecks created in ports when lines that comprise many containers are held until all of the line has been cleared. Container mode makes it much quicker for border staff to manage entries with a large number of items, but means that when the data are analysed, entries in container mode are censored --- in these cases, it is unknown which items were inspected, and which of those were found to be compliant or non-compliant. This makes analysis of the data to identify trends in biosecurity risk challenging. 

 Data that have been collected in container mode are closely related to the data collected under \emph{pooled testing}, which is often used for disease surveillance. Within the pooled testing literature there are two main branches: one aims to identify positives within a pool, while the other seeks to use pooled data to estimate quantities about the population. It is the latter -- estimating quantities -- that we are interested in. The fundamental problem is estimating a prevalence, \(p\), within a population, when only pooled data are available \citep{thompson_estimation_1962}. More recent work has focused on improving estimates by reducing bias, either through altering the sampling strategy \citep{schaarschmidt_experimental_2007,hepworth_debiased_2009} or by incorporating bias correction into models  \citep{hepworth_bias_2017,hepworth_bias_2021}. There have also been extensions of the problem where \(p\) is not a constant, but it is estimated using linear regression using only the pooled data \citep{delaigle_nonparametric_2015, chatterjee_regression_2020, mcmahan_bayesian_2017, liu_generalized_2020}. These papers have made significant progress in fitting increasingly complex models, but do not focus on the impacts of different types of pooling on the precision of model estimates.

In practice, however, pooled testing is manifestly different from the biosecurity scenario. Pooled testing exists by design: as a way to gather information about a population while reducing testing. In biosecurity, inspection is applied to every individual line, and the results are only pooled at the point of data capture --- \rev{as a side effect of the mode selection of the entry}. Hence, we are interested in how much information we are losing due to aggregating results as container mode. \rev{Our analysis offers regulators the opportunity to assess the risks of the continued use of container mode, and to weigh them against its operational advantages.}

In this paper we investigate the effect of container mode data collection upon our ability to estimate the biosecurity risk of items. We start with an asymptotic analysis, where we calculate the precision of estimates and determine the implications of mixing different item types in container mode. We then develop a simulation experiment that allows us to understand how larger entries and more item types affect the precision of our estimates. Finally, we analyse biosecurity data provided by the Australian Department of Agriculture, Fisheries and Forestry (DAFF) to identify the real-world differences between using the line-only data and including the container mode data.
\section{Model overview}
\subsection{Data}\label{sec:data}

To make our language about the data more precise, we will explicitly define what we mean by entries, lines and directions. Entries are a collection of lines, and a line is a group of the same type of item or commodity being imported. When cargo enters the country, each line is given \emph{directions}. These directions detail all of the activities undertaken by the biosecurity regulator to manage the biosecurity risk of each line, and also the outcome of those activities. \rev{To isolate the uncertainty that arises from use of container mode, we focus on one aspect of the directions recorded against import lines: whether they were deemed compliant (within biosecurity regulations) or not.}

\rev{Line and container mode are the two ways that records are kept of actions made by the biosecurity regulator.} In line mode, the directions assigned to each line are recorded along with the outcome for that line. In container mode, directions are only recorded per entry. This means that in container mode, if any line in that entry has an inspection, and non-compliance is found, then every line in that entry is recorded to be non-compliant. If all of the lines in the entry are compliant, then they are all marked compliant, so in this case line and container mode are equivalent.

\subsection{Modelling}\label{sec:model}

Throughout this paper, we focus on estimating the probability that a line is non-compliant using information including the type of item, country of origin and whether it has complete documentation. The full model for the probability that a line is non-compliant, \(p_{ijk\ell}\), is: 

\begin{align}
\logit(p_{ijk\ell}) = \alpha_{i} + \beta\mathbb{I}_j + \delta_k + \gamma_\ell \label{eq:logit_model_sim},
\end{align}
where the fixed effects are $\alpha_i$, $\beta\mathbb{I}_j$ and $\delta_k$: $\alpha_i$ represents the item type, the indicator variable $\mathbb{I}_j$ denotes whether there is correct documentation, the coefficient $\beta$ is the given to the instance that there is correct documentation, and $\delta_k$ represents the country of provenance. The random effect $\gamma_\ell$ represents the entry effect, which we include because there may be correlations between lines within an entry. \rev{We anticipate some correlation because lines in the same entry originate from a common context, so they may be non-compliant for related reasons}. The indices can take values

\begin{align}
i &=1, \ldots, a, & a&\in \mathbb{N},& a &= \text{\# items}\\
j &= 1, 2, & & & &\text{\hspace{-5pt}without and with documentation}&\\
k &= 1,\ldots, d,& d &\in \mathbb{N},& d &= \text{\# countries}\\
\ell &= 1,\ldots,g,& g&\in \mathbb{N},& g &= \text{\# entries}.
\end{align}
The values of the indicator variable are

\begin{align}
\mathbb{I}_1 & = 0,& &\text{without documentation}\\
\mathbb{I}_2 &= 1,& &\text{with documentation}.
\end{align}
The random effect $\gamma_\ell$ has distribution

\begin{align}
\gamma_\ell | \sigma &\sim \Normal(0, \sigma), & \ell &= 1,\ldots, g, & g&\in \mathbb{N}.
\label{eq:entry_effect}
\end{align}

If all data were in line mode then the above model would be a fairly standard mixed effects logistic regression with categorical variables. However, because of the use of container mode to capture the data, we don't observe outcomes for each line, as every line in the entry is marked as non-compliant if any line in the entry is found to be non-compliant. Therefore, the outcome is whether the entry is compliant and we need to calculate the probability that the entry is non-compliant, which is one minus the probability that every line in the entry is compliant:

\begin{align}
\mathbb{P}\left(\text{Entry } l \text{ non-compliant}\right) = q_l = 1-\prod_{ijk \text{ for lines in }l}[1-p_{ijkl}],\label{eq:entry_pr}
\end{align}
where \rev{$p_{ijkl}$ is the probability that the line with indices $ijkl$} is non-compliant, calculated from Eq.~\eqref{eq:logit_model_sim}.
Hence, for entries in container mode, we treat the entry  as a Bernoulli random variable with probability defined by Eq.~\eqref{eq:entry_pr}, while for entries in line mode, we treat each line as a Bernoulli random variable with probability as defined in Eq.~\eqref{eq:logit_model_sim}.

This paper includes three analyses: an asymptotic analysis,
 a simulation study, and a case study of Australian biosecurity data. \rev{For the asymptotic analysis we only consider the item type, ignoring effects due to the country of origin, documentation and entry effect. As such,} rather than using Eq.~\eqref{eq:logit_model_sim}, we just consider the probability that a line of item type \(i\) is non-compliant, \(p_i\). The simulation study and the case study both use the full model, as defined above. 

\section{Asymptotic analysis}\label{sec:asymptotic_analysis}
We use asymptotic analysis to investigate how the precision of estimates depends on entry size, the number of entries, the probability of non-compliance and whether item types are mixed. This analysis comprises two parts. The first assumes that all items are a single type, which allows us to quantify how the amount of data, probability of non-compliance  and entry size affect precision. The second part assumes that there are two different item types, and it explores how changing the proportion of entries with both item types mixed affects precision.

Throughout this section we make two simplifications. Firstly, we do not separate line mode and container mode because container mode data with an entry size of one is mathematically equivalent to line mode data. Hence, throughout these analysis, an entry size of one means line mode and entry size greater than one implies container mode. Secondly, we  assume that each item has a fixed probability of non-compliance. \rev{As such, any uncertainty in the inference of this value arises as a result of the difference between container and line mode. With this in hand,} we consider each line a Bernoulli trial, which only depends on the item type. When the entry size is greater than one, the relevant probability is whether at least one line was \rev{non-compliant.}

\rev{We estimate precision using an asymptotic estimate of the standard error. We calculate  the precision from the square roots of the diagonal elements in the Fisher information matrix, $\mathcal{I}$, which is the expected value of the negative of the Hessian matrix of the log-likelihood evaluated at the value of the parameter.}


\subsection{Single item type}
For the single item type case, we set the probability of non-compliance to be \(p\), and define \(N\) as the total number of entries, \(I\) as the number of \rev{non-compliant entries} and \(S\) as the size (i.e. the number of lines) in each entry. The likelihood is a binomial distribution, where the outcome is \rev{the discovery of a non-compliant entry}. The probability that an entry is \rev{compliant} is

\begin{align}
\mathbb{P}(\text{entry compliant}) = (1-p)^S,
\end{align}
meaning that the binomial likelihood for a set entry size $S$ \rev{proportional to}

\begin{align}
\mathcal{L}_S = \left(1-(1-p)^S\right)^I\left(1-p)^S\right)^{N-I}. \label{eq:entry_mode_like}
\end{align}
Therefore, the log-likelihood is

\begin{align}
  \log\mathcal{L}_S& = {I} \log\left(1-(1-p)^S\right)+(N-I)\log\left((1-p)^S\right).
  \label{eq:loglikelihood_arbitrary_entry_size}
\end{align}
\ifproofread
\textbf{\hl{Edited: omitted generalisation to varying entry sizes.}}
\fi
As there is only one parameter, we calculate its second derivative (rather than needing a Hessian matrix):

\begin{align}
\left[\frac{\partial^2 \log\mathcal{L}_S}{\partial p^2}\right]= 
\frac{S\left(N + \frac{I((1+S)(1-p)^S-1)}
{\left((1-p)^S-1 \right)^2} \right)}
{(1-p)^2}.\label{eq:loglike_2nd_deriv}
\end{align}
To calculate the Fisher information, we take the expectation of the number of entries with non-compliance, which depends on the size of the entry:

\begin{align}
\mathbb{E}\left[I\right] = N(1-(1-p)^S). \label{eq:exp_intercept}
\end{align} 
Hence the Fisher information is

\begin{align}
\mathcal{I} = -\mathbb{E}\left[\frac{\partial^2 \log\mathcal{L}_S}{\partial p^2}\right]=
 -\frac{NS^2(1-p)^{S-2}}{(1-p)^S-1}, \label{eq:fisher_information}
\end{align}
and the standard error estimate is

\begin{align}
SE = \left(-\frac{NS^2(1-p)^{S-2}}{(1-p)^S-1}\right)^{-1/2}. \label{eq:general_SE_calc}
\end{align}

\begin{figure}[h!]
\begin{subfigure}[b]{0.49\textwidth}
\includegraphics[width=\textwidth]{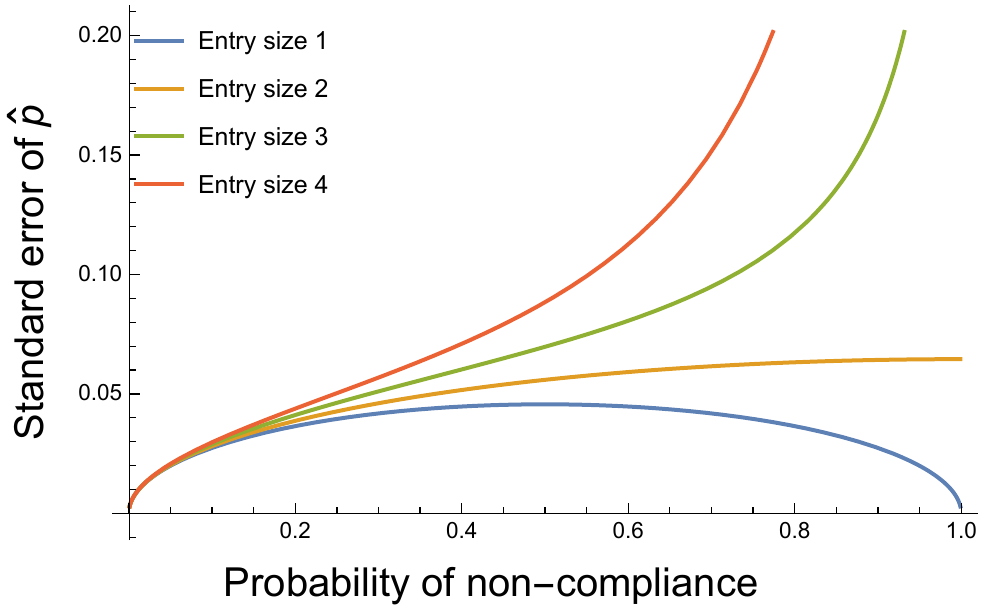}
\end{subfigure}
\hfill
\begin{subfigure}[b]{0.49\textwidth}
\includegraphics[width=\textwidth]{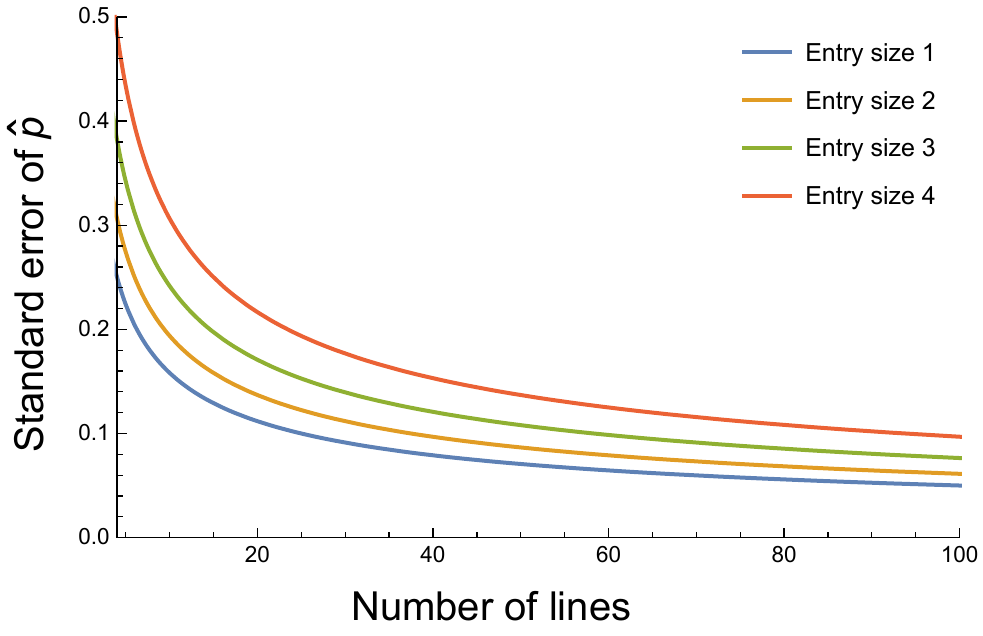}
\end{subfigure}

\caption{The standard error (Eq.~\eqref{eq:general_SE_calc}) as the probability of non-compliance, \(p\), is varied (left) and as the number of lines are varied (right). For the left plot the number of lines is held constant at 120. For the right plot, the probability of non-compliance is held constant at 0.5}
\label{fig:asymptotic_SE_plots}
\end{figure}

Using Eq.~\eqref{eq:general_SE_calc} we can understand how the probability of non-compliance, the entry size and the number of lines affect the standard error, and we plot these relationships in Figure~\ref{fig:asymptotic_SE_plots}. The left plot shows that the standard error depends on the probability of non-compliance and that the relationship depends on the entry size. For all entry sizes, the standard error is small when the probability of non-compliance is small (below \(\sim 0.3\)). However, for larger values of the probability of non-compliance, the standard error increase significantly if the entry size is three or greater. The large \(p\) behaviour is driven by the \((1-p)^{S-2}\) term in Eq.~\eqref{eq:general_SE_calc}, which means SE goes to 0 if \(S=1\), while it diverges if \(S\geq 3\). Figure~\ref{fig:asymptotic_SE_plots} also shows how the standard error decreases as the number of lines of data increases. The lower the entry size is, the lower the standard error, and, as \(SE\sim\sqrt{1/N_{E,S}}\), container mode data with larger entry sizes require a large amount of data to reach the same standard error.

\subsection{Two item types}
In this section we consider a situation where there are two items with probabilities of non-compliance of \(p_1\) and \(p_2\), and we examine how these different items probabilities interact. We focus on a scenario where every entry is of size two, meaning there are three types of entries: only type 1; only type 2; or mixed, with one line of type 1 and one of  type 2. We denote the number of lines within a single entry of type 1 and 2 as \(S_1\) and \(S_2\) respectively, and \(I(S_1,S_2\) and \(N(S_1,S_2)\) are the number of entries with non-compliance and total number of entries with \(S_1\) type 1 lines and \(S_2\) type 2 lines. For our case, we can have \(S_1=2, S_2=0\); \(S_1=1, S_2=1\); or \(S_1=0, S_2=2\). Rewriting the log-likelihood from Eq.~\eqref{eq:loglikelihood_arbitrary_entry_size}, we get

\begin{align}
\log\mathcal{L} =& \sum_{S_1,S_2}{I(S_1,S_2)}\log\left(1-(1-p_1)^{S_1}(1-p_2)^{S_2}\right) + \nonumber
\\ &\quad\quad(N(S_1,S_2)-I(S_1,S_2))\log\left((1-p_1)^{S_1}(1-p_2)^{S_2}\right). \label{eq:entry_mode_loglike_2types}
\end{align}
We compute the Fisher information matrix and the standard error using Mathematica. The standard error for \(p_1\) is

\begin{align}
\frac{1 }
{2}
\sqrt{
-\frac{\textstyle{{p_1} \left(p_1^2-3 p_1+2\right) (N_{1,2} (p_1-1) (p_2-2) p_2}{+4 N_{2,2} (p_2-1) (p_1 (p_2-1)-p_2))}}
{\splitfrac{N_{1,1} (p_1-1) (N_{1,2} (p_1-1) (p_2-2) p_2+4 N_{2,2} (p_2-1) (p_1 (p_2-1)-p_2))}{+N_{1,2} N_{2,2} (p_1-2) p_1 (p_2-1)^2}}
}
,\label{eq:p1_two_type_SE_est}
\end{align}
where \(N_{1,1}\) and \(N_{2,2}\) are the number of entries with two type 1 lines and two type 2 lines respectively, while \(N_{1,2}\) are the number of entries with both type 1 and type 2 lines. The standard error for \(p_2\) is the same, with \(N_{1,1}\) and \(N_{2,2}\) switched and \(p_1\) and \(p_2\) switched. 

By examining Eq.~\eqref{eq:p1_two_type_SE_est} we can see that the behaviour of the standard errors is more complex than when we considered only one type of item. Notably, the number of entries of only type 2, \(N_{2,2}\), is in the equation for the type 1 standard error, along with the probability of non-compliance of type 2, \(p_2\). 

\begin{figure}[H]
\begin{subfigure}[b]{.49\textwidth}
\includegraphics[width=\textwidth]{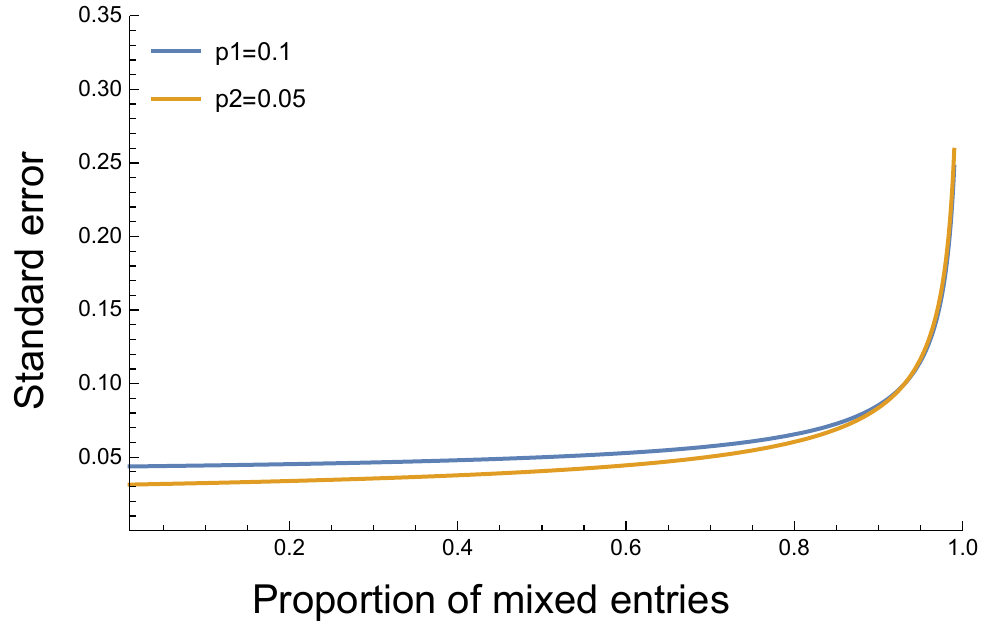}
\end{subfigure}
\hfill
\begin{subfigure}[b]{0.49\textwidth}
\includegraphics[width=\textwidth]{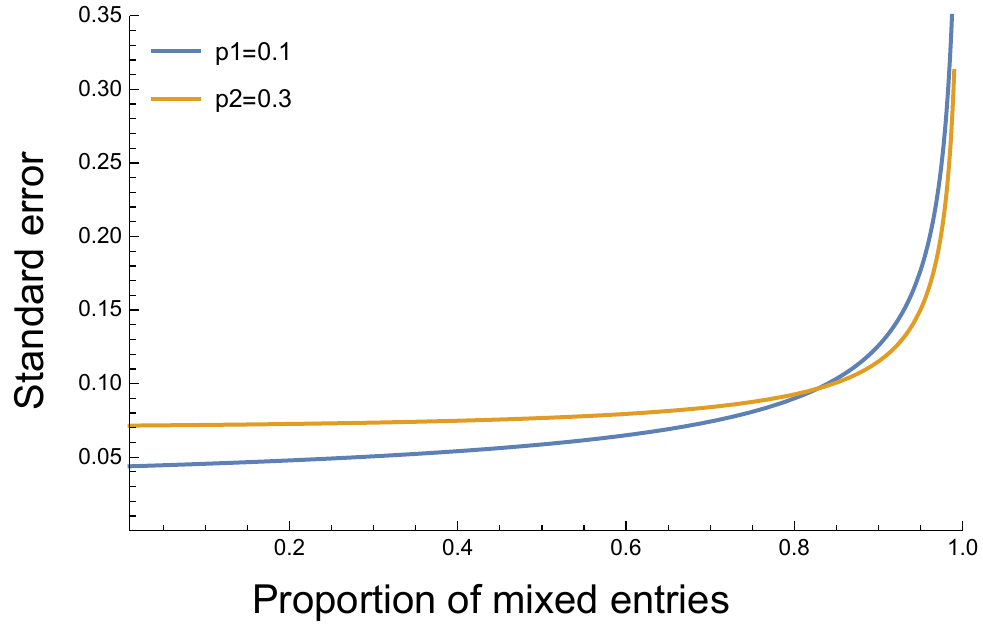}
\end{subfigure}
\hfill
\begin{subfigure}[b]{0.49\textwidth}
\includegraphics[width=\textwidth]{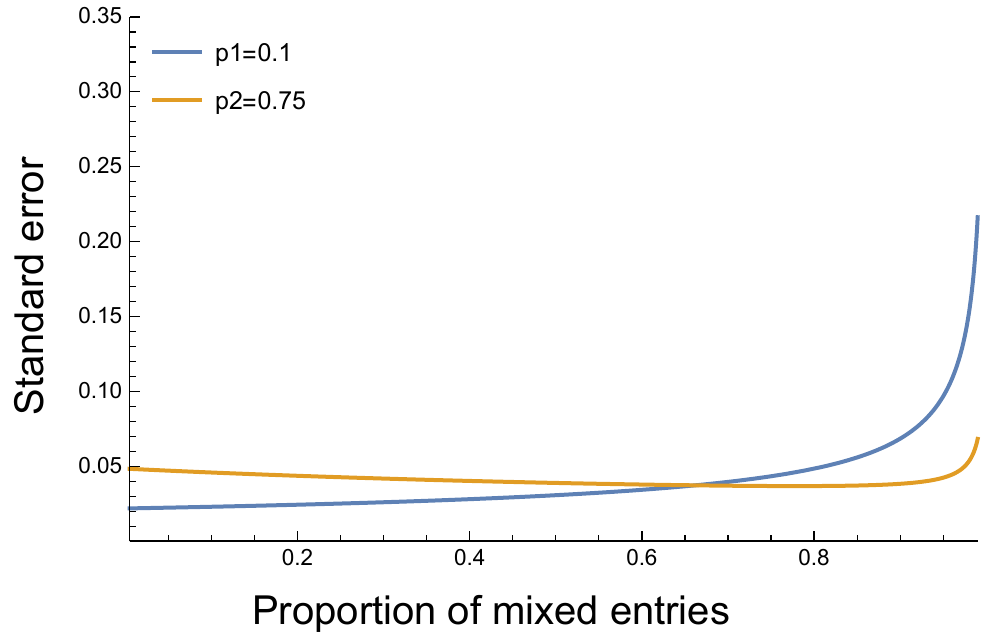}
\end{subfigure}
\begin{subfigure}[b]{0.49\textwidth}
\includegraphics[width=\textwidth]{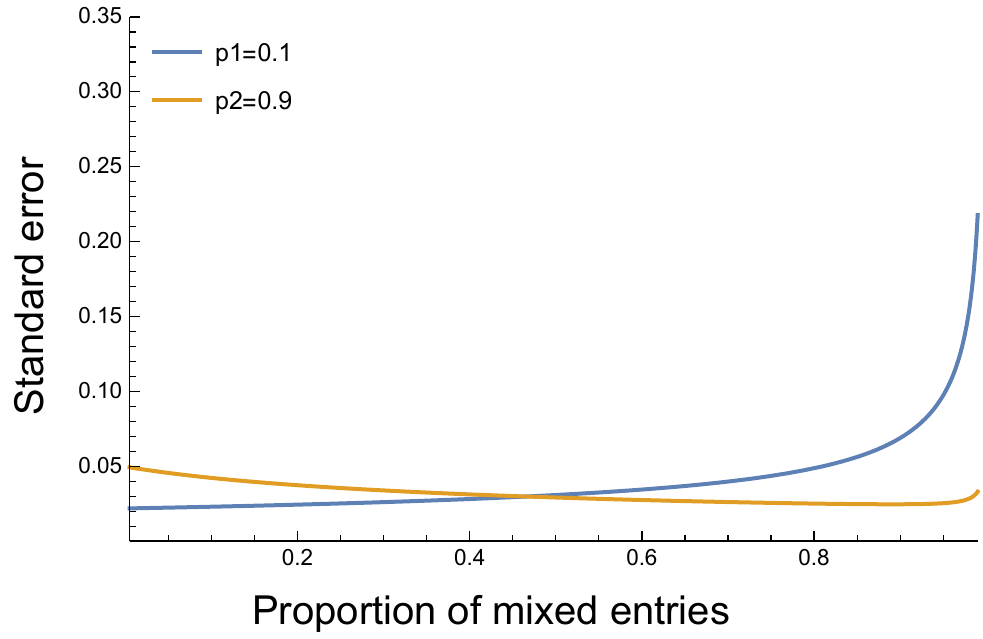}
\end{subfigure}

\caption{The standard error when for entry size 2, when the proportion of mixed entries is varied. In each plot there are 50 total entries and \(p_1\) is held constant at 0.1, while \(p_2\) goes from 0.05 up to 0.9.}
\label{fig:two_type_se_fourplot}
\end{figure}

Figure~\ref{fig:two_type_se_fourplot} shows how the standard error varies with the proportion of mixed entries, for different non-compliance probabilities. \rev{Here we fix $N_\text{total}=50$, and keep $N_{1,1}=N_{2,2}$ while the proportion $N_{1,2}/N_\text{total}$ is varied.} In each case, we hold \(p_1\) at 0.1, and we vary \(p_2\) for 0.05 up to 0.9 across the four plots. The way that the standard error changes as a function of the proportion of mixed entries changes markedly, depending on the value of \(p_2\). In particular, when \(p_2\) is 0.7 and 0.9, the standard error for \(p_1\) increases as the proportion of mixed entries increases, while the standard error for \(p_2\) actually decreases, while the proportion of mixed entries is below \(\sim 90 \%\).

%
%
%
%

\subsection{Asymptotic analysis conclusions}

From the asymptotic analysis we make two observations:
\begin{enumerate}
\item Container mode data are most useful when the probability of non-compliance, \(p\), is small. In the extreme case where \(p\rightarrow 0\), container mode data are equivalent to line mode data because whenever an entry is found to \rev{be compliant, we know that every line within that entry is compliant}, regardless of whether it is container mode or line mode. 
\item \rev{By allowing entries to contain lines of different item types, container mode} can increase or decrease the precision of estimates, depending on the true probability of non-compliance.
\end{enumerate}

\section{Simulation study}
As the first step in developing a method to analyse real-world data sets, we simulate a data set that contains key complexities including multiple item types, fixed effects and random effects. Once we have simulated data, we can fit a model to estimate parameters. The advantage of our approach is that we can (1) verify that our model behaves correctly and (2) explore how model precision varies in a more complex setting.

\subsection{Data simulation}
We simulate data from the full model (as described in Section~\ref{sec:model}). For our simulations, we chose parameters that we expect to be similar to real-world values. We set \(\beta=-1\), choosing an negative effect of having correct documentation, because we expect lines with correct documentation have a higher chance of being compliant. We include $a=5$ item types with \(\alpha_i\) taking values of -6.91, -4.60, -3.89, -2.94, -1.39 (corresponding to non-compliance probabilities of 0.001, 0.01, 0.02, 0.05 and 0.2, if all other effects were 0). We chose negative values for the item effect $\alpha_i$, since the probability of detection is expected to be low for any item in real-world data. We used $d=3$ countries and set their weights to be 0.5, -1 and 0.25.
For within-entry correlation, we set \(\sigma = 0.25\). For each entry we draw whether it is in line mode or not with probability 0.25, and for each line we set the probability of having correct documentation to be 0.2. We show an example of the format of the data in Table \ref{table:example_data}.

\vspace{0.1cm}
\begin{table}[h]
\caption{Example simulated data. Lines 1-3 are all marked as non-compliant because entry 1 is in container mode, even though only 1 or 2 of the lines were actually non-compliant.}
\label{table:example_data}
\begin{center}

\begin{tabular}{|c|c|c|c|c|c|}
\hline 
Line & Entry & Type & Documentation & Mode & Non-compliant \\ 
\hline 
1 & 1 & 3 & 1  & Container & 1 \\ 
\hline 
2 & 1 & 2 & 0  & Container & 1 \\ 
\hline 
3 & 1 & 1 & 0 & Container & 1 \\ 
\hline 
4 & 2 & 5 & 0 &Line & 0 \\ 
\hline 
5 & 3 & 4 & 1 &Line & 1 \\ 
\hline 
\vdots & \vdots & \vdots & \vdots & \vdots & \vdots \\ 
\hline 
\end{tabular} 

\end{center}
\end{table}

\subsection{STAN model}
We model the system in a Bayesian framework, using the RSTAN package in R. We choose a Bayesian package due to the ease of specifying the model and could be updated, in principle, for any specification of the model for the probability of non-compliance. The basic structure for fitting the simulated data is to define the probability of non-compliance for each line (following Eq.~\eqref{eq:logit_model_sim})

\begin{align}
\logit(p_{ijk\ell}) = \alpha_{i} + \beta\mathbb{I}_j + \delta_k + \gamma_\ell, \label{eq:logit_model_fit_sim}
\end{align}
where 

\begin{align}
\gamma_\ell | \sigma &\sim \Normal(0, \sigma), & \ell &= 1,\ldots, g, & g&\in \mathbb{N}.
\end{align}
As we fit our model in a Bayesian framework, we set priors on each parameter:

\begin{align}
\alpha_i &\sim \Normal(-4, 4), & i&=1, \ldots, a, & a&\in \mathbb{N},\\
\beta &\sim \Normal(0, 0.5),\\
\delta_k &\sim \Normal(0, 0.5), & k &= 1,\ldots, d, & d &\in \mathbb{N}\\
\sigma &\sim \Normal(0, 0.5)
\end{align}

The probability $p_{ijk\ell}$ is the probability of non-compliance of a line that has characteristics $(i,j,k,\ell)$. We denote the probability that line $n$ is non-compliant as $p_{(n)}$, where $p_{(n)} = p_{ijk\ell}$ if that line has characteristics $(i,j,k,\ell)$.

We fit the STAN model to the simulated data for two reasons. Firstly we confirm that, with sufficient data, the STAN model can accurately estimate model parameters. Secondly we explore how changing the amount of data and the proportion of container mode data affects precision, over a range of entry sizes. To measure precision, we use the standard deviation of the posterior samples, because it is the Bayesian equivalent of the standard error calculations in Section~\ref{sec:asymptotic_analysis}.

\subsection{Simulation results}

\begin{figure}[h!]
\includegraphics[width=\textwidth]{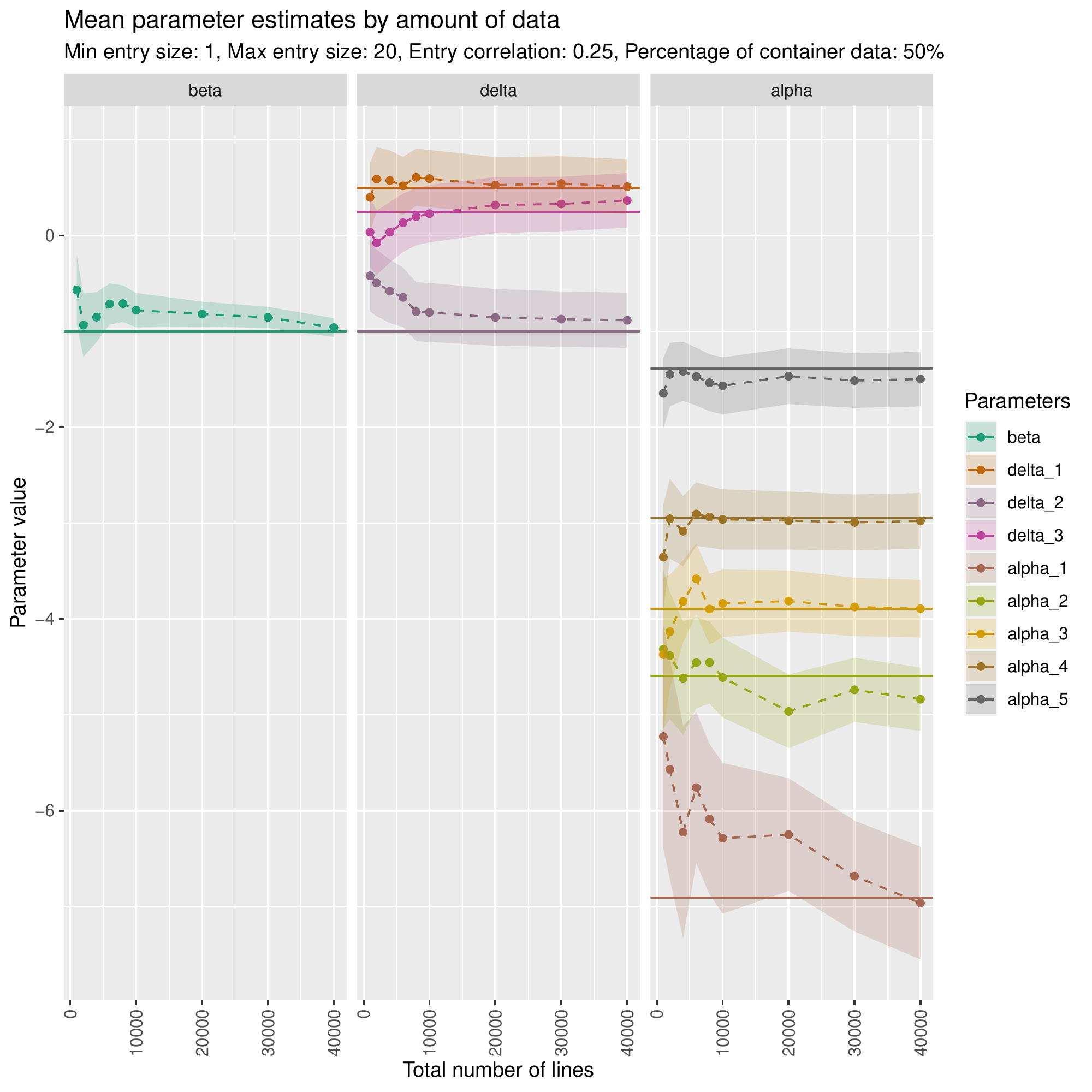}
\caption{Results from the STAN model, fitted to simulated data. 50\% of the data are in container mode and 50\% are in line mode, and each entry is a random size between 1 and 20. Dots and dashed lines in each plot show the STAN estimates, while the solid lines show the true value of each parameter.}
\label{fig:simulation_estimation}
\end{figure}

Figure~\ref{fig:simulation_estimation} shows the that the STAN model gives good parameter estimates, if there is sufficient data. Overall, the STAN estimates are close to the true values when there is a large amount of data. The two smallest values of \(\alpha\) show the worst performance, which is not surprising, given our asymptotic analysis results (Section~\ref{sec:asymptotic_analysis}).  

\begin{figure}[h!]
\includegraphics[width=\textwidth]{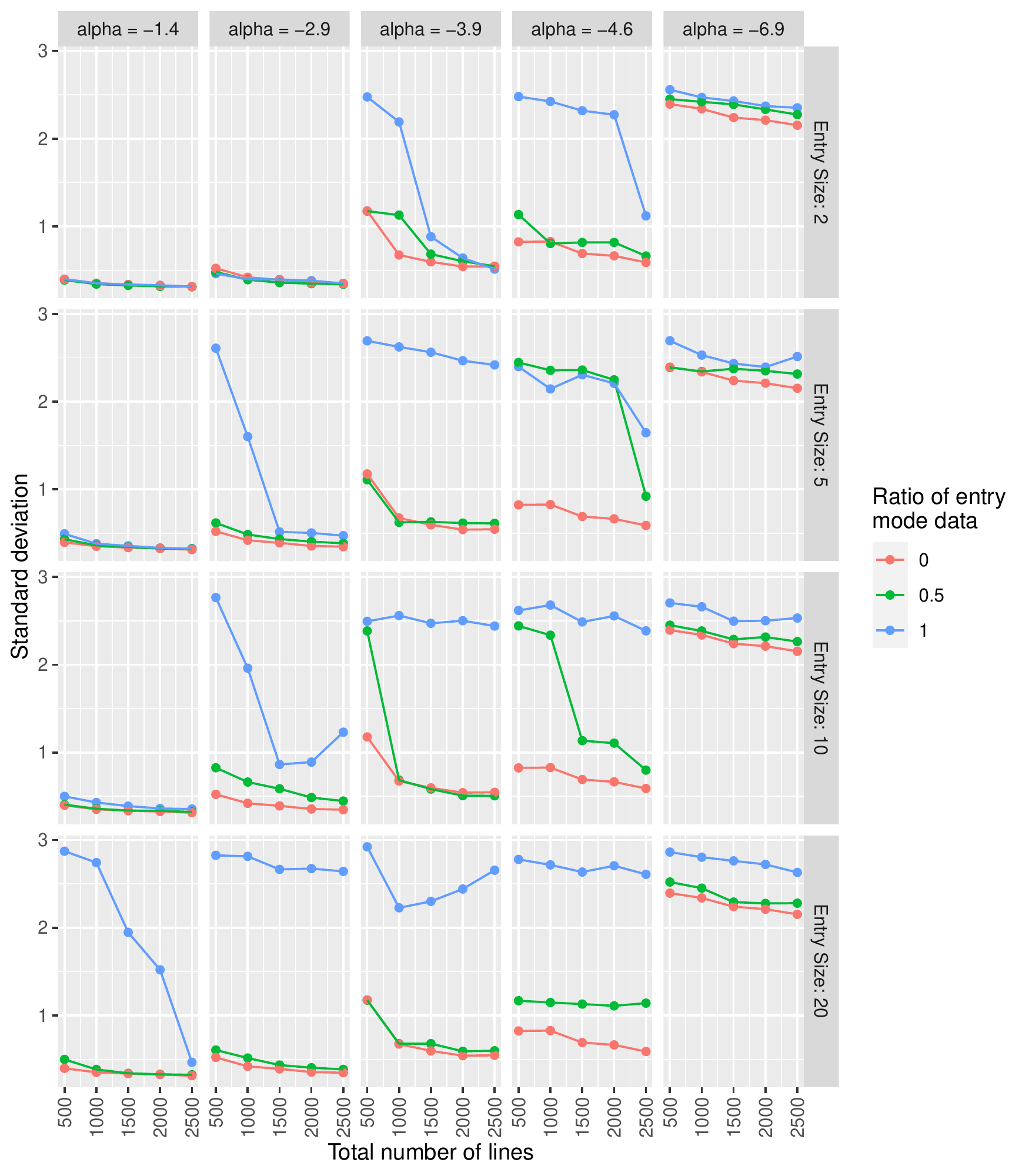}
\caption{The standard deviation of the alpha parameter estimates for simulations with varying entry size and total number of lines. The ratio of container mode data is the proportion of data in container mode, meaning that 0 corresponds to line mode data-sets, 0.5 if a mixed data-set and 1 is purely container mode. }
\label{fig:SD_est_simulated}
\end{figure}

The precision of model estimates depends on the entry size, the probability of non-compliance and the amount of container mode data (Figure~\ref{fig:SD_est_simulated}). The best-case scenario is when all data are in line mode (ratio of container mode = 0), and the distance from that line shows the impact of using a mix, or only, container mode data. We find that the difference between the precision estimates depends strongly on the combination of factors. There are some combinations where all analyses return similar precision (e.g. \(\alpha=-1.4\) and entry size 10), while other combinations have a large gap where the line only data far outperforms the others (e.g. \(\alpha=-4.6\) and entry size = 5).

\section{Case study}
We apply our model to some real biosecurity data, to see how our understanding of the system changes by fitting the full data, compared to only using the container mode data. The data set is of furniture imports in 2020, and we break the data set into 52 weekly data sets for this analysis. Because container mode data with no non-compliance is equivalent to line mode data, the more container-mode entries with non-compliance, the larger the differences could be between analysing all of the data compared to only the line mode data. There is a natural break in the dataset, where three weeks have four container mode entries with non-compliance, with the remaining weeks having fewer. Hence, we choose to analyse these three weeks as a demonstration of potential real-world differences between analysing all data and only line mode data. For confidentiality, for each week we re-name the item type, the country and the week to be integers (week 1, 2 and 3 do not correspond to the first 3 weeks of the year, and country 1 and item 1 are not the same in weeks 1 and 2). Relabeling the data does not cause issues because we are not trying to compare estimates between weeks or draw inferences between countries or item types in this analysis. A summary of the data is given in Table~\ref{tab:real_data_summary}.

\begin{table}[ht]
\caption{The number of lines in container mode and line mode for each week. The ratio is the fraction of the week's data that is in container mode.}
\label{tab:real_data_summary}
\centering
\begin{tabular}{rrrrr}
  \hline
 Week & Entry & Line & Ratio  \\ 
  \hline
 1 & 516 & 6741 & 0.07 \\ 
  2 & 1667 & 7213 & 0.19 \\ 
   3 & 679 & 6832 & 0.09 \\ 
   \hline
\end{tabular}
\end{table}

\begin{figure}[h!]
\includegraphics[width=\textwidth]{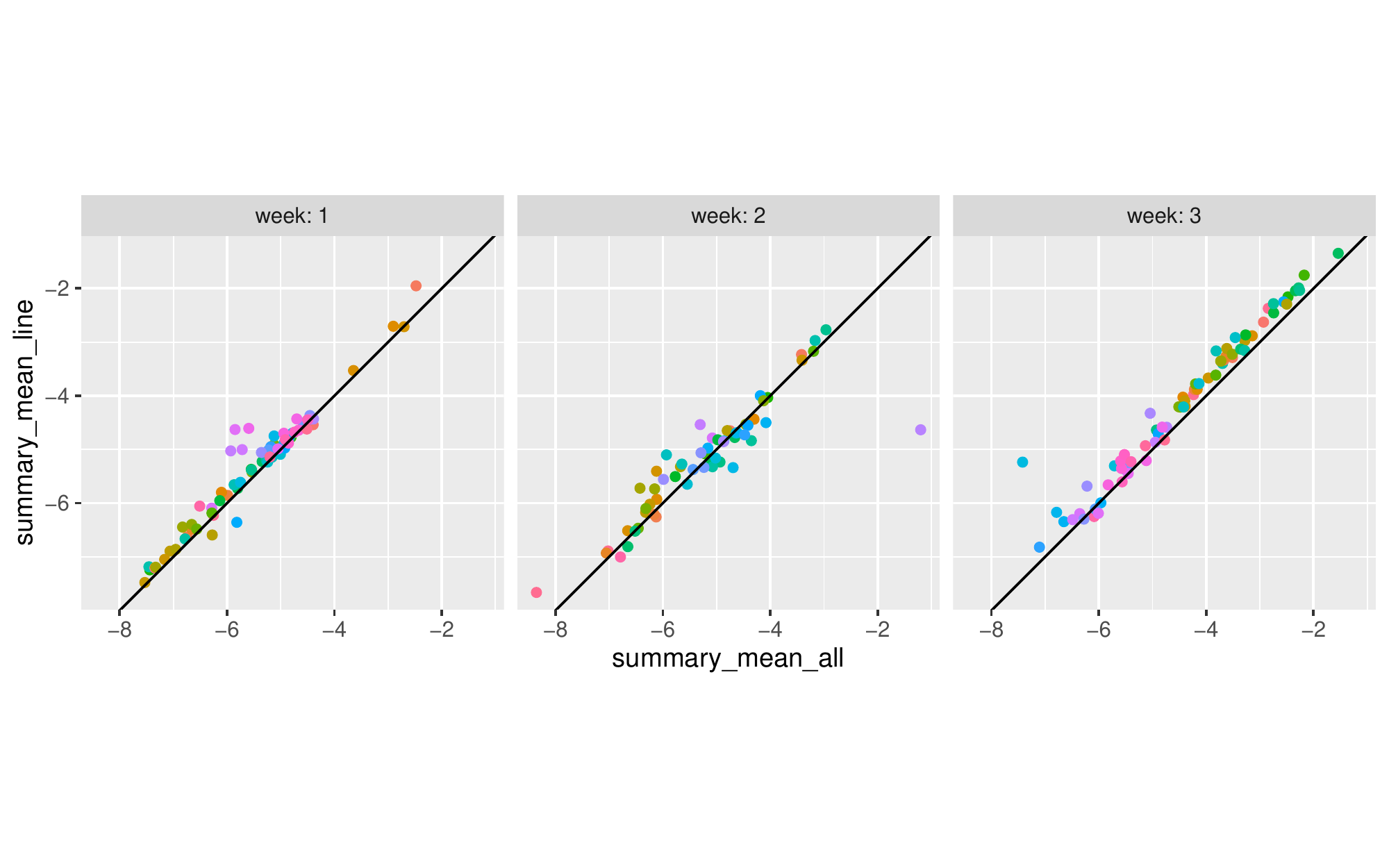}
\caption{Scatter plot of parameter estimates using the line only data and all data. Dots that fall on the solid black line indicate when we get the same parameter estimate from each data set.}
\label{fig:case_study_scatter}
\end{figure}

We fit our STAN model to the data from each week, using all the data and the line-only subset of the data. Because the line-only analysis is a subset of the data, not every country or item is present in the line-only data set. For the parameters that do match, we generate a scatter plot of the mean estimates to see how frequently we get different estimates (Figure~\ref{fig:case_study_scatter}). For many of the parameters, we get very similar results irrespective of which data we use. This is not surprising, given that most of the data are in line mode and that line mode data gives more information than container mode data. However, it is interesting that including the container mode data results in quite large changes to some of the parameter estimates. 

Analysing the full data set, rather than the line-only data, gives us information about more items and countries. Table~\ref{tab:estimates_only_from_all_data} shows all of the parameter estimates that we only get when including the container mode data in the analysis.

\begin{table}[ht]
\caption{Parameter estimates that are only possible when using all data, as opposed to the line-only fits. \rev{The value summary\_rhat\_all measures convergence of MCMC chains} \citep{brooks2011handbook}.}
\label{tab:estimates_only_from_all_data}
\centering
\begin{tabular}{rlrrrr}
  \hline
 & param\_name & week & summary\_mean\_all & summary\_sd\_all & summary\_rhat\_all \\ 
  \hline
1 & p\_intercept[72] & 1.00 & -1.70 & 1.68 & 1.00 \\ 
  2 & p\_intercept[59] & 2.00 & -1.34 & 1.71 & 1.01 \\ 
  3 & p\_intercept[60] & 2.00 & -2.07 & 1.63 & 1.04 \\ 
  4 & p\_intercept[61] & 2.00 & -1.18 & 1.92 & 1.00 \\ 
  5 & country\_effect[62] & 2.00 & -0.01 & 0.49 & 1.00 \\ 
  6 & country\_effect[63] & 2.00 & -0.01 & 0.47 & 1.00 \\ 
  7 & country\_effect[64] & 2.00 & -0.01 & 0.49 & 1.01 \\ 
  8 & country\_effect[65] & 2.00 & 0.02 & 0.48 & 1.02 \\ 
  9 & country\_effect[66] & 2.00 & 0.01 & 0.48 & 1.00 \\ 
  10 & country\_effect[67] & 2.00 & -0.01 & 0.48 & 1.00 \\ 
  11 & p\_intercept[72] & 3.00 & -1.43 & 1.78 & 1.00 \\ 
   \hline
\end{tabular}
\end{table}

\section{Discussion}

In this paper we sought to understand how container mode data affects our ability to estimate risks in the biosecurity setting. Due to the relative ease of analysing line data, in our experience, container mode data are often excluded from data analysis. From our analysis of real biosecurity data, we find that including the container mode data in analysis can markedly change some results and that by only using line data the analysis is somewhat limited -- there are parameter estimates that simply cannot be made. 

Through our simulation study and our asymptotic analysis we gain an understanding of how mixing container mode data into line mode data impacts our ability to make precise inferences. We see that mixing lines that have different probabilities of being non-compliant has divergent affects on model precision. In the asymptotic analysis we see that mixing a high-probability item with a lower probability item makes it easier to precisely estimate the high-probability item's parameter, while mixing two low-probability items makes it challenging to make inferences about either. \rev{When one item has a low probability of non-compliance, a non-compliant mixed entry is likely to correspond to a non-compliant entry of the of the item with a higher probability of compliance.  The extra data from this entry can therefore be used to refine the parameter estimates, whereas if the respective probabilities are not well resolved this is not possible.} These interactions between items are exacerbated in the simulation study, where we have larger entries and more item types, and we see differing patterns of item types, entry sizes and amounts of data where the presence of container mode data degrades model precision, compared to a line-only analysis. 

While we are able to model the full dataset, the continuing presence of container mode data will be a barrier for future data analysis. Most statistical and machine learning algorithms are designed such that each row of data will have explanatory variables and an outcome. With container mode data, we have the explanatory variables, but only partial information about the outcome. Hence, we need customised algorithms (such as described in this paper) to analyse it appropriately. \rev{Even so, many simplifying assumptions have been made: for example, that a given item's probability of non-compliance does not vary over time. These simplifications do not change our conclusions, as the analysis still serves to highlight the differences between container and line mode. However, without proper care, the specific results of our case-study analysis should not be applied to decision making.} 

While in our case the pooling of outcomes in container mode is an artifact of the system, pooled testing are often designed strategically in public health to improve efficiency. The strain on the PCR testing system has prompted regimes to identify cases while minimising the rounds of testing~\citep{mutesa_pooled_2021}. While it is clear that the way pooling is conducted affects how quickly infections can be identified, \rev{in some circumstances, the outcomes of pooled testing can also be used to estimate parameters, and our work has implications for these situations} \citep{delaigle_nonparametric_2015, chatterjee_regression_2020, mcmahan_bayesian_2017, liu_generalized_2020}. From our asymptotic analysis, it is clear that even when only estimating a prevalence, for a fixed number of tests, the standard error of the estimate depends on pool size and therefore there would be an optimal pool size, which would depend on the prevalence  \citep{keeling2022modeling}. When there is a model with parameters being estimated, the added complexity could compound the issue, making pool sizing a more important aspect of study design. Furthermore, as we demonstrate here, it is not only pool size that matters, it is how samples with different characteristics are grouped together. Hence, for a given study, there may not be one optimal pooling strategy, and there would likely be trade-offs between different parameters when it comes to maximising the precision of estimates. 

\rev{In this paper we have investigated how pooled data can reduce our ability to make statistical inferences about the population. Within the Australian biosecurity context, the pooling is due to how data are recorded and was implemented for operational reasons. Due to this operational decision, analysing full datasets is harder than the line-only mode data, meaning that either analysis is restricted to a subset of possible methods or that container mode data is ignored. While it does not follow that container mode should be removed due to operational efficiency, if the line-level data could be captured then it would improve our understanding of biosecurity risk. 
However, pooled data will continue to be collected in various fields, for example due to the cost-savings of testing multiple samples at once for diseases.
While there is extensive work in pooled testing protocols for case identification, there is less work on identifying pool sizes when aiming to make inferences about aspects of the population.
Our work shows that pooling tests with variable underlying prevalence affects precision differentially. Hence, careful considering should be given to designing pools when the data will be used to make inferences about the population.}

\section{Code availability}
All code is available at \url{https://github.com/cmbaker00/container-line-analysis-public}.

\section{Acknowledgements}
We would like to thank Tom Waring for their contribution to editing this manuscript.

\bibliography{container-line}{}
\end{document}